\documentclass{elsart}
\usepackage{graphicx}
\usepackage{amssymb}

\newcommand{\bes}{\ensuremath{^{7}}Be}
\newcommand{\bet}{\ensuremath{^{10}}Be}

\newcommand{\bel}{\ensuremath{^{11}}B}

\newcommand{\ctw}{\ensuremath{^{12}}C}
\newcommand{\cft}{\ensuremath{^{14}}C}

\newcommand{\helion}{\ensuremath{^3}He}

\newcommand{\ecm}{\ensuremath{E_\textrm{c.m.}}}
\newcommand{\thetacm}{\ensuremath{\theta_\textrm{c.m.}}}
\newcommand{\thetalab}{\ensuremath{\theta_\textrm{lab}}}
\newcommand{\eex}{\ensuremath{E_\textrm{ex}}}
\newcommand{\jpi}{\ensuremath{J^\pi}}

\newcommand{\degree}{\(^{\circ} \) \kern-.4em} 

\begin{document}

\begin{frontmatter}



\title{Experimental investigation of a linear-chain structure in the nucleus $^{14}$C}


\author[CNS]{H.~Yamaguchi\corauthref{cor1}}
\ead{yamag@cns.s.u-tokyo.ac.jp}
\author[CNS,Edinburgh]{D.~Kahl} 
\author[CNS]{S.~Hayakawa}
\author[CNS]{Y.~Sakaguchi}
\author[CNS]{K. Abe}
\author[CNS,JAEA]{T.~Nakao}
\author[Matsue]{T.~Suhara}
\author[Tohoku]{N.~Iwasa}   
\author[SKKU,Ewha]{A.~Kim}
\author[Ewha]{D.H.~Kim}
\author[SKKU]{S.M.~Cha}
\author[SKKU]{M.S.~Kwag}
\author[SKKU]{J.H.~Lee}
\author[SKKU]{E.J.~Lee}
\author[SKKU]{K.Y.~Chae}
\author[RIKEN]{Y.~Wakabayashi}
\author[CNS]{N. Imai}
\author[CNS]{N. Kitamura}
\author[Chungang]{P.~Lee}
\author[IBS,WNSC]{J.Y.~Moon} 
\author[IBS]{K.B. Lee}
\author[IBS]{C. Akers}
\author[WNSC]{H.S. Jung}
\author[IOP,Dongnai]{N.N. Duy}
\author[IOP]{L.H. Khiem}
and
\author[Chungang]{C.S.~Lee}

\corauth[cor1]{Corresponding author.}

\address[CNS]{Center for Nuclear Study (CNS), University of Tokyo, RIKEN campus, 
  2-1 Hirosawa, Wako, Saitama 351-0198, Japan}
\address[Edinburgh]{School of Physics and Astronomy, the University of Edinburgh,  Peter Guthrie Tait Road, Edinburgh EH9 3BF, UK}
\address[JAEA]{Advanced Science Research Center, Japan Atomic Energy Agency, Tokai, Ibaraki 319-1195, Japan}
\address[Matsue]{Matsue College of Technology, Matsue, Shimane 690-8518, Japan}

\address[Tohoku]{Department of Physics, Tohoku University, Aoba, Sendai, Miyagi 980-8578, Japan}

\address[SKKU]{Department of Physics, Sungkyunkwan University, 2066 Seobu-ro, Jangan-gu, Suwon, Korea}
\address[Ewha]{Department of Physics, Ewha Womans University, Seoul 120-750, Korea}

\address[RIKEN]{ The Institute of Physical and Chemical Research (RIKEN), 
2-1 Hirosawa, Wako, Saitama 351-0198, Japan}

\address[Chungang]{Department of Physics, Chung-Ang University, Seoul 156-756, Korea}

\address[IBS]{Institute for Basic Science, 70, Yuseong-daero 1689-gil, Yuseong-gu, Daejeon 305-811, Korea}

\address[WNSC]{Wako Nuclear Science Center (WNSC), KEK, 2-1 Hirosawa, Wako, Saitama 351-0198, Japan}

\address[IOP]{Institute of Physics, Vietnam Academy of Science and Technology, 10 Dao Tan, Ba Dinh, Ha Noi, Vietnam}

\address[Dongnai]{Dong Nai University, Le Quy Don Street, Tan Hiep Ward, Bien Hoa City, Dong Nai, Vietnam}

%\author{L.H.~Khiem\\
%Institute of Physics and Electronics, Vietnam Academy of Science and Technology\\
%1}

\begin{abstract}
It is a well-known fact that a cluster of nucleons can be formed in 
the interior of an atomic nucleus, and such clusters
may occupy molecular-like orbitals, showing 
characteristics similar to normal molecules consisting of atoms.
Chemical molecules having a linear alignment
are commonly seen in nature, such as carbon dioxide.
A similar linear alignment of the nuclear clusters,
referred to as linear-chain cluster state (LCCS), has been  
studied since the 1950s, however, up to now there is no
clear experimental evidence demonstrating the existence of such a state.
%% The studies were mainly concerned with light 4$n$-nuclei (e.g. $^{12}$C and $^{16}$O),
%% which are comprised of multiple $\alpha$ particles.
%% The failure to discover the LCCS may indicate the difficulty of the creation of 
%% an LCCS in a system comprised of only $\alpha$-particle clusters, 
%% and a special mechanism may be required to stabilize it.
Recently, it was proposed that an excess of neutrons may offer just such 
a stabilizing mechanism, revitalizing interest in the nuclear LCCS, 
specifically with predictions for their emergence in neutron-rich carbon isotopes.
Here we present the experimental observation of $\alpha$-cluster
states in the radioactive $^{14}$C nucleus. %, which is of great importance
%also for the radiometric dating method.
Using the \bet+$\alpha$ resonant scattering
method with a radioactive beam, we observed a series of levels 
which completely agree 
with theoretically predicted levels having an 
explicit linear-chain cluster configuration.
We regard this as 
the first strong indication of the linear-chain clustered nucleus.
%This work can be considered as an important first step as a new technique 
%in the alignment and morphology of 
%the atomic nucleus, which could be built upon in the future to enable the 
%production and study of nuclei with more exotic topology.
\end{abstract}

\end{frontmatter}

% main text
%\section{Introduction}
%\label{sec:intro}

Atomic nuclei are frequently observed to manifest effects of a clustered 
substructure within, and
the particular importance of $\alpha$ particle clustering  
was pointed out even in the earliest works of nuclear physics \cite{Bethe:36,Hafstad:38}.
%\cite{Bethe:36,Wefelmeiyer:37}.
%cite Bethe & Bacher (1936), Bertsch and Bertozzi
%Clustering in atomic nucleus is a common phenomenon
%$\alpha$-cluster states are known to emerge
%in many nuclei, including isotopes of the carbon nucleus.
In 1956, Morinaga \cite{Morinaga:56} came up with the novel idea of 
a particular cluster state: the linear-chain cluster state (LCCS).
In that work, it was suggested that the 7.66-MeV state in \ctw\
--which is now known as the Hoyle state-- 
may correspond to a state of three $\alpha$ particles arranged in a row.
Similar $\alpha$-clustering in other 4$n$-nuclei, which are  
comprised of multiple $\alpha$ particles,
was also discussed in the same work.
Later, it was shown by Horiuchi \cite{Horiuchi:75}
%using the orthogonality condition model (OCM)
that the Hoyle state could be a molecular-like level of $^8$Be+$\alpha$,
or equivalently three $\alpha$ particles weakly coupled to each other, 
instead of an LCCS. 
However, the concept of the LCCS has particularly drawn 
the attention of nuclear physicists, both experimentally and theoretically.
Now the LCCS is commonly considered as extreme and exotic, 
due to its presumed propensity to exhibit bending configurations.
Therefore, its identification would have a strong impact 
on the research field of quantum many-body systems.

Despite its pursuit by many scientists for more than half a century,
up until now the LCCS has been only hypothetical.
%Among the $\alpha$-cluster states, linear-chain cluster states
%have been considered as exotic and of interest for a long time. 
There have been LCCS candidates, such as the one in $^{16}$O 
proposed by Chevallier {\it et al.} \cite{Chevallier:67}
based on the large moment of inertia found for 
an assumed rotational band. 
However, the spin-parity assignment of the 19.3-MeV level 
was questioned in later experiments \cite{Freer:95,Curtis:13}, 
and the interpretation as an LCCS was not supported
by recent theoretical works \cite{Ohkubo:10,Ichikawa:11,Suhara:14}.
%% The search for linear-chain cluster structure in carbon or other nuclei 
%% has a long history.
As for the carbon isotopes,
Itagaki {\it et al.} \cite{Itagaki:01} studied
$\alpha$-cluster states in $^{12,14,16}$C using a microscopic model. 
They investigated breathing and bending motions and
concluded that $^{16}$C might have a linear-chain structure 
%with a stability 
but at high excitation energies \eex\ $>$ 20~MeV.
%In their following work \cite{Itagaki:04}, it was discussed that 
%an equilateral-triangular shaped cluster structure 
%is expected to emerge in \cft.  
%Such a structure should be reflected to 
%a negative parity rotational band with the $K$-number $K = 3$,
%the first members of which may correspond to the known states at 
%$\eex = 9.80$ MeV (spin and parity $\jpi=3^-$) and 11.67 MeV (4$^-$).
In the work by Oertzen {\it et al.} \cite{Oertzen:04}, 
it was proposed that prolate deformed %$K^{\pi}=0^{\pm}$ 
bands should exist in \cft.
Their idea was that those bands might be attributed to 
an underlying LCCS structure,
but the reasoning was merely based on the relatively 
large momentum of inertia, and 
the spin and parity \jpi\ were confirmed only for low-lying levels %(0$^+$, 1$^-$, and  2$^+$) 
in the bands.
The LCCS in $^{13}$C have also been studied both experimentally and 
theoretically \cite{Milin:02,Itagaki:06,Suhara:11,Yoshida:13,Yamada:15}; however,
%theoretically \cite{Milin:02,Itagaki:06,Suhara:11,Yamada:15}; however,
there is no agreed-upon interpretation that 
the observed cluster levels may arise from an LCCS.
In summary, there is no clear evidence of any LCCS in nuclei at present.

\begin{figure}[!htbp]
\centerline{\includegraphics{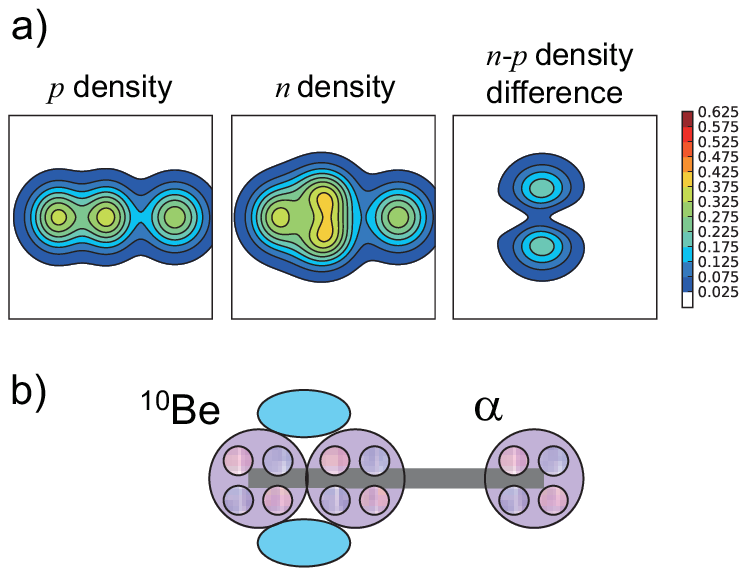}}
\caption{\label{fig:cluster} 
Wavefunction dominant in the LCCS in \cft\ calculated by the AMD method \cite{Suhara:10a,Suhara:11}.
a) Proton density $\rho_p$, neutron density $\rho_n$ and the difference between them. 
The box size is 10 $\times$ 10 fm$^2$ for all.
b) An intuitive picture of the above wavefunction.}
\end{figure}

A theoretical prediction of LCCS in \cft\
was made by Suhara and En'yo \cite{Suhara:10a}
with an 
antisymmetrized molecular dynamics (AMD) calculation, 
yielding a prolate band (\jpi\ = 0$^+$, 2$^+$, 4$^+$) 
that has a configuration of an LCCS
at a few MeV or more above the 
\bet+$\alpha$ threshold. %unlike the prolate bands in the study of Oertzen {\it et al.} \cite{Oertzen:04}.
They showed that the LCCS is stabilized by its 
orthogonality to lower-lying states.
At lower excitation energy in $^{14}$C, 
there are triaxially deformed cluster states, which are constructed 
by bases with bending configurations.
To fulfill the orthogonality condition between different states, 
higher-excited LCCSes are prohibited from bending.
This is in stark contrast with $^{12}$C, where no triaxial 
bands exist, and therefore such an LCCS-stabilizing mechanism does not work.
A further investigation \cite{Suhara:11} showed that the AMD wavefunction
has a configuration in which two $\alpha$ particles and two neutrons 
are located close to each other, 
while the remaining $\alpha$ particle is 
relatively further away, as illustrated in Fig.~\ref{fig:cluster}.
This implied that such an LCCS could be experimentally 
accessible from the \bet+$\alpha$ channel in a single step.
%%%----new
The emergence of the prolate band as LCCSes had been 
discussed in the previous pioneering work \cite{Oertzen:04},
and two essential new features found in \cite{Suhara:10a} are
the absence of the negative-parity band, which appears to be contradictory to  
the concept of the parity inversion doublets \cite{Butler:96}, and 
the higher level energies above the \bet+$\alpha$ threshold. 
The former was explained as the result of a stronger mixing of 
the negative parity LCCS, in which the \bet\ core can be rotated easily, 
with other bending-shaped configurations.
The latter can result from the consumption of kinetic energy 
from the linear-chain alignment.
%They claim the triaxial bands obtained at the sub-threshold energy 
%may correspond to the $K=0^+$ rotational band in \cite{Oertzen:04}.

%%%----new

\begin{figure}[!htbp]
\centerline{\includegraphics{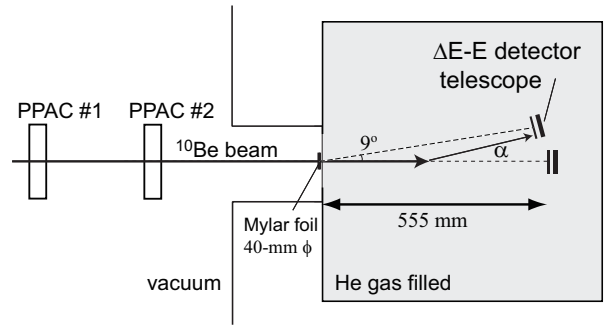}}
\caption{\label{fig:setup} 
The experimental setup for the resonant scattering measurement. 
}
\end{figure}

The excited states in \cft\ have been studied by
various reactions  \cite{Oertzen:04,TUNL14:91,Resler:89,Soic:03,Milin:04,Price:06,Price:07,Haigh:08,Cappuzzello:15},
%such as $^9$Be($^6$Li, \proton)\cft\  \cite{TUNL14:91,TUNL}.
%Some recent studies were motivated by the interest on the cluster structure in \cft,
%and performed  with
%%*Soi\'c {\it et al.},
%$^7$Li($^9$Be, \cft$^*$)$d$ \cite{Soic:03},
%2$n$-transfer   \cite{Oertzen:04}, 
%*Milin {\it et al.}, 
%reactions of \ctw($^6$He, $\alpha$)\cft$^*$ \cite{Milin:04},
%\cft($^{13}$C, \cft$^*$) \cite{Price:06},
%\cft(\cft, \cft$^*$) \cite{Price:07},
%and \ctw($^{16}$O, $^{14}$O)\cft$^*$  \cite{Haigh:08},
%where \cft$^*$ could be detected as \bet+$\alpha$ or $^{13}$C+$n$.
%There are resonances
%commonly observed in several methods,
but only the excitation energies are known for most levels.
%which never lead us to a conclusive confirmation of 
%the LCCS.
In the present work we applied the \bet+$\alpha$ resonant scattering method 
in inverse kinematics \cite{Artemov:90} to identify 
the predicted LCCS band in \cft.
Our experimental setup was similar to the previous one 
in the \bes+$\alpha$ experiment \cite{Yamaguchi:13}, 
but we placed an extra silicon detector telescope
to cover a broader angular range, instead of the NaI detectors,
as shown in Fig.~\ref{fig:setup}.
%and we used the PPACs as the beam counter.
The new setup enabled us to perform a reliable analysis on the angular distribution.
%with a better position sensing of the beam and recoiled particles.
An advantage of the present method is that 
only natural parity levels ($\pi=(-1)^{J}$) % 1$^{-}$, 2$^{+}$, ...)
are selectively observed since both particles have $\jpi=0^{+}$.
The coverage of the most forward laboratory angle \thetalab, 
corresponding to the 
center-of-mass angle $\thetacm= 180$\degree,
provided us with the clearest identification of the resonances,
because the Coulomb potential scattering is at minimum there,
%Another advantage of the present angular range is that 
and it suffers the least from the 
uncertainty of the nuclear phase shift.
%which is often approximated as a hard-sphere phase shift 
%in R-matrix analyses. %calculations on the determination of the resonant parameters.

Two similar measurements have been independently planned, 
carried out and published recently. 
The first work by Freer {\it et al.} \cite{Freer:14} 
had a similar setup to ours, but with a more limited % 
angular sensitivity. 
%but the present work is concentrated more on the LCCS
%predicted at lower energies, and a more comprehensible 
%analysis including angular dependence has been performed.
Another work by Fritsch {\it et al.} \cite{Fritsch:16}
used an active target setup, but detection was only possible for side scattering angles.
%but the beam statistics was one order of magnitude smaller than us.

%The cross section was measured with
%the thick-target method in inverse kinematics \cite{Artemov:90}.
The present measurement was performed at the low-energy 
radioactive isotope beam separator CRIB \cite{Kubono:02,Yanagisawa:05,Yamaguchi:08}.
The \bet\ beam was produced 
% \cite{Kubono:02,Yanagisawa:05,Yamaguchi:08}.
%The \bet\ beam was produced  
via the \bel($^2$H, \helion)\bet\ reaction
in inverse kinematics using a 1.2-mg/cm$^2$-thick deuterium gas target
and a \bel\ beam at 5.0 MeV/u accelerated with an AVF cyclotron.
%The \bet\ beam was separated and purified by magnetic analysis and velocity 
%selection with a Wien filter.
The \bet\ beam had a typical intensity of
2 $\times$ 10$^4$ particles per second, and the 
beam purity was better than 95\%.
%at the gas target of scattering measurement.
The beam was counted with two parallel-plate avalanche counters (PPACs), 
% \cite{Kumagai:01}, 
%separated by 30~cm,
which enabled us to perform an unambiguous event-by-event beam particle identification 
with the time-of-flight information.
%The PPACs measured the timing and position of the incoming \bet\ beam ions
%with a position resolution better than 1 mm. 
%The trajectory of each beam particle was reconstructed with 
%an extrapolation of the positions measured with the PPACs. 
%The timing signal was used for the beam particle identification 
%using the time-of-flight method.
The \bet\ beam at 25.8 MeV impinged on
the gas target, which was a chamber filled with
helium gas at 700~Torr (930 mbar) and 
covered with a 20-$\mu$m-thick Mylar film
as the beam entrance window.
The measured \bet\ beam energy at the entrance of the 
helium gas target, after the Mylar film, was 24.9$\pm$0.3~MeV.
$\alpha$ particles recoiling to the forward angles
were detected by $\Delta E$-$E$ detector telescopes.
We used two sets of detector telescopes in the gas-filled chamber,
where each telescope consisted of two layers of 
silicon detectors with the thicknesses of 20~$\mu$m and 480~$\mu$m.
The central telescope was located 555 mm downstream of the beam entrance window
exactly on the beam axis,
and the other telescope was at an angle of 9\degree\ from the beam axis, 
as viewed from the entrance window position.
%The helium gas was sufficiently thick to stop the 
%\bes\ beam in it before reaching the central telescope.
Each detector in the telescope had an active area of 50 $\times$ 50~mm$^2$,
and 16 strips for one side, 
making pixels of 3 $\times$ 3~mm$^2$ altogether. 
These detectors were calibrated with $\alpha$ sources,
as well as with 
$\alpha$ beams at various energies produced 
during the run. 
%Each detector had an energy resolution better than 1.5\%
%in full width at half maximum (FWHM)
%for 5-MeV $\alpha$ particles.
%The beam energy was degraded in the thick gas target and 
%recoiling $\alpha$ particles originating from 
%elastic scatterings reached the telescope.
The main measurement using the helium-gas target 
was performed for 2 days, injecting 
2.2 $\times 10^{9}$ \bet\ particles into the gas target as valid events.
%% We performed the same measurement but using an argon-gas target 
%% of an equivalent thickness for 1 day in order to 
%% evaluate the background $\alpha$ particles 
%% directly reaching the telescopes as a contamination in 
%% the incoming secondary beam or other sources of beam-induced background.  
%% The predominant species measured at the telescopes
%% were $\alpha$ particles, which were distinguished from other particles
%% using the $\Delta E$-$E$ information. % from the elastic scattering.

We selected genuine scattering events based on the coincidence of a \bet\ particle 
incident on the target, as determined from PPAC trajectory and time-of-flight measurements, 
with an $\alpha$ particle incident on the silicon detectors.
%We selected events corresponding to a  \bet\ beam particle injected into 
%the target for which an $\alpha$ particle was detected at the silicon detector telescope 
%in coincidence with the \bet\ at the PPAC.
A precise energy loss function of the \bet\ beam in the %Mylar window and the  
helium gas target was obtained by a direct energy measurement at seven different target pressures
interpolated with a calculation using the SRIM \cite{Ziegler:08} code.
The scattering position, or equivalently the 
center-of-mass energy \ecm, 
was determined by a kinematic reconstruction on an event-by-event basis.
%which uses the energy loss function, the beam trajectory measured 
%with the PPACs, and 
%the energy and position of the recoiled $\alpha$ particle at the telescope.
The number of events for each small energy division was 
converted to the differential cross section ($d\sigma/d\Omega$)$_\textrm{c.m.}$, 
using the solid angle of the detector, the number of beam particles, 
and the effective target thickness, without any artificial scaling.
%% The number of beam particles was precisely known by the single counting of the 
%% beam ions with the PPAC, simultaneously recorded in the measurement.
%% The background contribution, evaluated by the argon-target run data,
%% was subtracted from the helium-target spectrum.
Finally we obtained the excitation function
of the \bet+$\alpha$ resonant elastic scattering for 13.8--19.1 MeV, % as shown in Fig.~\ref{fig:ex_functions}a,
where events with $\thetalab=$ 0--8\degree\ 
(\thetacm = 164--180\degree) 
were selected.
The overall uncertainty in \ecm\ was estimated as
80--110~keV, depending on the energy.
The uncertainty mainly originated from 
the energy straggling of the \bet\ and $\alpha$ particles (40--50~keV) and
the energy resolution of the detector telescopes (50--100~keV).
%% The angular uncertainty was about 1.5\degree\ in \thetalab, arising from
%% the precision of the detector alignment, the finite size of the pixel of the telescopes,
%% and the extrapolation of the PPAC positions.
%where  \bes$^*$ or \bes$^*$ denotes the first excited state of each nucleus.
%The data in the lowest energy region ($\eex < 14.2$ MeV) have a larger uncertainty 
%because in those events the $\alpha$ particle was stopping
%in the first layer of the telescope,
%and the position resolution was limited.
%The particle identification may not be complete there.

\begin{figure}[!htbp]
\centerline{\includegraphics{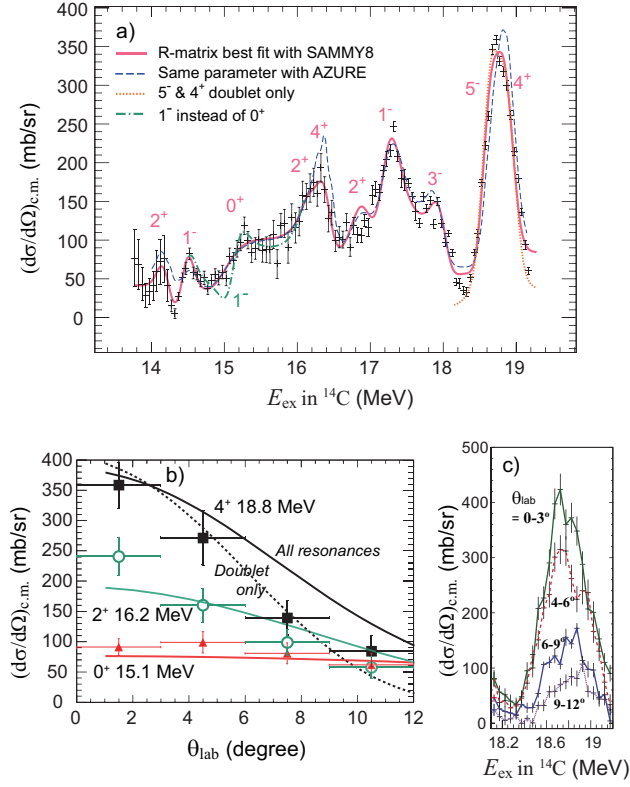}}
\caption{\label{fig:ex_functions} 
Experimental center-of-mass cross section of the \bet+$\alpha$ resonant scattering ($d\sigma/d\Omega$)$_\textrm{c.m}$.
a) Excitation function  for $\thetalab=$ 0--8\degree,
fitted with R-matrix calculations;
the best R-matrix fit with SAMMY8 \cite{SAMMY:00} (solid curve, $\chi^2$/$n_\textrm{dof}=62/82$),
the same fit but with the 0$^+$ resonance replaced with 1$^-$ (dash-dotted curve), 
and another fit only with a doublet of 4$^+$ and 5$^-$ around 18.7~MeV (dotted curve).
The dashed curve is with the 
above best fit parameters but calculated with AZURE.
b) Angular distribution of ($d\sigma/d\Omega$)$_\textrm{c.m.}$ at the resonant energies of the 
0$^+$, 2$^+$ and 4$^+$ levels. The present experimental data points are compared with
the R-matrix calculations, drawn as the curves.
%The abscissa is the scattering angle of the recoiled $\alpha$ particle
%in the laboratory frame \thetalab.
c) Peak profiles around 18.7~MeV for several angular ranges.
}
\end{figure}

The elastic-scattering excitation function 
we obtained is shown in Fig.~\ref{fig:ex_functions}a.
At energies above 15.7~MeV, the excitation function 
shows a reasonable agreement in the spectral shape 
with one of the recent measurements \cite{Freer:14}, 
although the previous absolute cross section appears to be larger by a factor of four.
We regard the difference as from an error in the overall 
normalization in the previous work, 
independent of the energy.
In fact, the previous analysis employed a 
normalization 
factor to adjust the absolute cross section,
%based on the Rutherford scattering cross section,
while in the present work the cross section was  deduced purely
from the experimental parameters, which is more reliable.
The overall agreement in the spectral shape
provides us a confirmation that there is no significant background contribution
induced by beam impurities or the inelastic channel,
because those depend significantly on the beam and 
target conditions in the setup,
which were quite different between the two measurements.
There is a larger disagreement in the lower energy region 
of 15.0--15.7~MeV, where a correct evaluation of the energy
loss function is essential.
The peak positions in the present measurement also resemble those of another
experiment with a break-up reaction \cite{Soic:03}.
Finally, the other elastic-scattering measurement \cite{Fritsch:16} yielded 
smaller cross sections, %(10--20 mb/sr at $\thetacm=$ 70--110\degree)
%% although 
%% we cannot directly compare it to the present work
%% due to the difference in the covered angular range.
%At their angle range, however,
%the cross section should be enhanced at the low energies,
%as the Rutherford cross section becomes 83 mb/sr at \ecm=2 MeV and \thetacm=90%\degree.
%% However, their cross section at low energy was 
even much smaller than the Rutherford cross section at low energies.
This is fundamental and contradictory to their large $\Gamma_\alpha$, but was unexplained.
Although they interpret their data as providing clear \jpi\ assignments,
their angular distributions show considerable deviation from the calculated distributions,
and the separation between the individual resonances was not clearly presented.
They claim they identified inelastic scattering events 
as a sharp locus in the correration plot of the scattering angles of \bet\ 
and $\alpha$, but a true inelastic scattering locus 
will exhibit a variable position depending on \ecm, and
the sharp locus never corresponded to the broad \ecm\ distribution they observed.
%%and most of the resonance energies are  
%% inconsistent with other measurements. 
Thus we do not employ their results as a credible source
in the present discussion.

We performed an R-matrix calculation % \cite{Lane:58} 
with SAMMY8 \cite{SAMMY:00}
to deduce the resonance parameters.
The energy broadening due to the experimental resolution was included 
in the R-matrix calculation,
and the \bet+$\alpha$ channel radius was taken to be 5.0~fm,
which was the distance obtained in the AMD calculation.
A deviation of $\pm0.5$ fm in the channel radius 
was accounted for in the systematic error. % of $\Gamma_\alpha$.
We also performed a calculation with AZURE \cite{AZURE} 
to evaluate the consistency between the calculation codes.
We confirmed the results are essentially consistent with one another, although 
minor differences are seen for closely spaced resonances.
%% which is the distance of the \bet\ and $\alpha$ particles 
%% obtained in the AMD calculation,
%% and also close to the value used in the other analysis \cite{Freer:14} (5.2 fm).
The main analysis was performed with a single channel, {\it i.e.}, 
only introducing the $\alpha$ particle decay width $\Gamma_\alpha$, 
which is a reasonable assumption to make when considering 
the basic characteristics of strong $\alpha$ resonances.
A multi-channel analysis introducing the $^{13}$C+$n$ channel 
was also performed, and the primary effect of the 
neutron channel was confirmed to be a simple reduction of the resonance height,
when the neutron width $\Gamma_n$ is comparable or larger than $\Gamma_\alpha$.
%% It implies that the neutron channel does not significantly affect the 
%% analysis of the energy and \jpi, 
%% but it may reduce $\Gamma_\alpha$ for resonances 
%% with large $\Gamma_n$, as discussed below.
%% We confirmed the R-matrix calculation in the previous work \cite{Freer:14} 
%% is reproduced with our code, which also proves
%% the absolute cross section is of the same definition 
%% between the two experiments.

The best fit parameters obtained from the analysis 
are summarized in Table~\ref{tab:resonance_params}.
%where we had a reasonable goodness of the fit as $\chi^2$/$n_\textrm{dof}=62/82$.
%The resonance parameters, in particular \jpi, 
%in the present energy region had scarcely been determined
%by previous experiments \cite{TUNL14:91, Soic:03,Oertzen:04,Milin:04,Price:06,Price:07,Haigh:08,Freer:14,Cappuzzello:15},
%and it was essential to determine \jpi\ independently from our own data.
Also shown in Table~\ref{tab:resonance_params} are 
the resonance parameters of the LCCS %of all the natural parity states 
obtained by the AMD calculation, where  
%and other previous experiments. 
the absolute level energies 
were normalized so that
the experimental \bet+$\alpha$ threshold energy at $\eex=$ 12.01~MeV is exactly reproduced. 
Such a normalization is known to 
provide a better reproducibility of experimental level energies
in the vicinity of the threshold, and 
we adopt the 
normalized \eex\ throughout this Letter. %in Table~\ref{tab:resonance_params} and the discussion below. 
%% The predicted LCCS had $\eex =$ 15.1, 16.0, and 19.2 MeV,
%% as listed in Table~\ref{tab:resonance_params}.
Resonances observed in previous experiments with unique determination of $\jpi$ 
are also listed (see \cite{Haigh:08} for a more complete tabulation, and 
\cite{Freer:14,Fritsch:16} for the lateset scattering experiments).
We do not find a clear correspondence for most resonances
because our measurement selectively observes natural-parity and $\alpha$-cluster-like states,
and is not very sensitive to the high-spin levels close to the \bet+$\alpha$ threshold. 
Here we describe the identification of the resonances around the predicted LCCS energies,
which forms the most essential part of the analysis.

1) {\it 15.1 MeV}; 
We observed relatively broad bumps around 14.5 and 15.1~MeV,
which are only consistent with low-spin resonances.
The best fit for these resonances are with $\jpi=1^{-}$ and 0$^{+}$, respectively.
If \jpi\ is assigned to be 1$^{-}$ for the latter, it 
significantly fails to reproduce the experimental cross sections 
at the lower energy side of the peak around 15~MeV, as shown as the dash-dotted 
curve in Fig.~\ref{fig:ex_functions}a.
Therefore, we adopt 0$^{+}$ as a unique assignment 
for the resonance around 15.1~MeV.
The angular distribution supports the assignment of 0$^{+}$ as well;
in Fig.~\ref{fig:ex_functions}b, the angular dependence 
of the experimental and calculated differential cross sections 
are compared.
% where \thetalab\ is 
%the angle of the recoiled $\alpha$ particle in the laboratory frame.
%The R-matrix calculation curves primarily resemble 
%the angular dependence of the squared Legendre polynomial, 
%but they are modified by the interference 
%with other resonances and the potential scattering.
The curve for 15.1~MeV is almost flat, being consistent with the 
\jpi\ = 0$^{+}$ assignment. % of the  resonance obtained above.  
%% Note that the fitting with these low-spin resonances would not be possible if
%% the cross section was four times more or larger, which may explain why
%% Freer {\it et al.} \cite{Freer:14} could not provide the R-matrix fit 
%% result for the low-energy region.

2) {\it 16.2 MeV}; 
A peak was observed around 16.2~MeV, and its tail on the 
low energy side was only consistent with a 2$^{+}$ resonance,
and it cannot be fitted with a resonance with  any other \jpi.
The sharp drop at the high energy side can be reproduced with 
another higher-spin resonance, and we introduced a 4$^{+}$ resonance
to obtain the best fit.
Once again, the angular distribution in Fig.~\ref{fig:ex_functions}b
is consistent with the above assignment, although not in perfect agreement, showing a  
modest change according to the angle.
The disagreement %%is still of a comparable order with the experimental uncertainty
%% shown in the plot, which 
%% includes the systematic error from the detector geometry and
%%background event subtraction as well as the statistical error,
%% and the observed deviation 
can be attributed to the interference with 
the nearby 4$^{+}$ resonance.

3) {\it 18.7 MeV}; 
Around 18.7~MeV we observed a strong peak that
diminishes quickly as \thetalab\ increases,
which is a clear indication of a higher spin ($J > 3$) state.
The width of the peak is twice as large as the experimental
resolution, and we could not obtain a satisfactory fit with only a single 
resonance, while the angular distribution is closest to that of a 4$^{+}$ resonance.
We obtained the best fit for this peak as a doublet of 4$^{+}$  and 5$^{-}$,
from all the possible combinations of two resonances.
A closer look at the resonance profiles at different angles  
is given in Fig.~\ref{fig:ex_functions}c, and one may notice that 
the centroid of the peak is increasing as \thetalab\ is increased.
This is consistent with the \jpi\ assignment, in which a lower-spin resonance 
is located at the higher-energy side.
In the previous work \cite{Freer:14}, the \jpi\ assignment of the resonance 
was 5$^{-}$  and they also introduced an additional 4$^{+}$ resonance,
but at the low-energy side.
%In the present work, the introduction of 
%4$^{+}$ resonance is essential to reproduce the 
%angular distribution.
We could not obtain a perfect fitting for the tails of this 
strong peak, namely around 18.2 and 19.2~MeV,
%we found an offset in the cross section 
%was created as 
due to an artifact of the R-matrix calculation
induced by the inclusion of low-spin resonances.
%This problem could be avoided by fitting the doublet 
%without any low-lying resonances, or reducing the channel radius.
The dotted curves in Fig.~\ref{fig:ex_functions}a and \ref{fig:ex_functions}b
are formed by fitting only with the doublet, which better reproduces the experimental data.
%and the agreement with the experiment was improved 
%at both sides of the peak. 
%The angular distribution is also better reproduced with the 
%doublet-only fit, as shown in Fig.~\ref{fig:ex_functions}b.
%The uncertainty of $\Gamma_\alpha$ was evaluated to cover both cases,
%fitting with all resonances, and with the doublet only. 

\begin{figure}[!htbp]
\centerline{\includegraphics{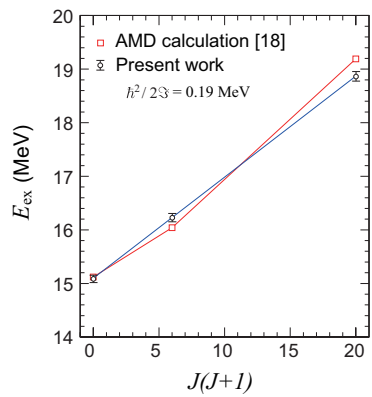}}
\caption{\label{fig:band} 
The $J(J+1)$-dependence of \eex\ for 
the band identified in the present experimental work, 
and the LCCS band predicted in  \cite{Suhara:10a}.
}
\end{figure}

%% As a result of the analysis, we identified three resonances around the 
%% energies of the predicted LCCS. The \jpi\ of those were uniquely determined from the resonant profiles 
%% as 0$^+$, 2$^+$, and 4$^+$,  
%% in consistent with the angular distribution 
%% as shown in Fig.~\ref{fig:ex_functions}b.
%% From an analysis on the 18.7-MeV peak profiles shown in Fig.\ref{fig:ex_functions}c,  
%% the 4$^+$ resonance is considered to comprise a doublet with a neigbouring 5$^{-}$ resonance.
%As a result, we identified three resonances around the 
%energies of the predicted LCCS. 
Although the aforementioned analysis was performed without 
any assumption from the theoretical calculation,
we identified three resonances %with unique \jpi\ assignments 
perfectly corresponded to the predicted LCCS band;
\jpi\ are identical,
and their energies and spacings are consistent with the theoretical prediction.
To illustrate the rotational feature of these levels,
the experimental and theoretical level energies are
plotted against $J(J+1)$ in Fig.~\ref{fig:band}.
It shows that both sets are almost on the same line,
$E_J=E_0+\hbar^2/2\Im (J(J+1))$,
where $\Im$ is the moment of inertia of the nucleus. 
The linearity allows us to interpret the levels  as a rotational band, and 
the low $\hbar^2/2\Im=$ 0.19~MeV implies the nucleus could be strongly deformed,
consistent with the interpretation of an LCCS.
%% There are non-LCCS states obtained in the AMD calculation \cite{Suhara:10a,Suhara:11}
%% as well, and their energies also exhibit a certain similarity with the experimental spectrum.
%% Nevertheless, we do not discuss their comparison in this Letter,
%% because all the non-LCCS states arose as predominantly neutron-excited levels,
%% whereas the present experiment is sensitive to $\alpha$-cluster-like states,
%% implying the similarity is likely to be accidental.
%% This can be contrasted with the LCCS,
%% which arise as strong-coupling cluster states with larger $\alpha$ widths,
%% providing us a reliable comparison with the present experiment.
%% These states refore, the observed resonances other than the above three
%% are considered to be general $\alpha$-cluster-like states that 
%% are not covered in the present AMD calculation.
%%% new
Although we observed several negative-parity resonances, 
we could not identify a negative-parity rotational band, 
which could be expected as the
counterpart of the parity inversion doublet.  
This shows that either the band is out of the sensitivity of our measurement,
or the negative-parity LCCSes are dissipated to other states as envisaged
by \cite{Suhara:10a}.
%%% new

The experimental $\Gamma_\alpha$ of these resonances %the LCCS candidates
are also compared with the theoretical predictions in terms of 
the dimensionless partial width $\theta_\alpha^2$
in Table~\ref{tab:resonance_params},  
although the precision of both is quite limited.
The experimental $\theta_\alpha^2$ was calculated as 
$\theta_\alpha^2 = \Gamma_\alpha / \Gamma_w$,
where $\Gamma_w$ is the Wigner limit of $\Gamma_\alpha$,
given by %$\Gamma_w=2 \hbar/R_n (2E/\mu)^{1/2} \mu P_l$ 
$\Gamma_w=\frac{2 \hbar^2}{\mu R^2} P_l$.
Here, $\mu$ is the reduced mass of the system and $P_l$ is the penetrability
of the $\alpha$ particle in the nucleus calculated with the interaction radius of $R=$ 5.0~fm,
for a given orbital angular momentum $l$.
%The overlap of the LCCS and $^{10}$Be+$\alpha$ channel wavefunctions
%was evaluated in a previous work with the AMD 
%calculation \cite{Suhara:10x},
It is not straightforward to obtain the theoretical $\theta_\alpha^2$ with an AMD calculation,
and in the present work, we calculated $\theta_\alpha^2$
with the method which evaluates the widths using
the overlap between the AMD and Brink wavefunctions \cite{Enyo:14}.

The calculation qualitatively reproduces the feature that the 
experimental $\theta_\alpha^2$ is anti-correlated with $J$.
This behavior corresponds to the reduction in the
overlap of the LCCS and the \bet(0$^+$)+$\alpha$ channel wavefunctions
for the higher-spin resonance found in the previous calculation \cite{Suhara:10x},
which can be explained with the mixed configuration of the LCCS.
%According to \cite{Suhara:10x},
The $4^+$ state is less mixed with the bending configurations in which the core \bet\ is rotating, 
resulting in a smaller overlap with the \bet(0$^+$)+$\alpha$ channel wavefunctions,
while the overlap with the \bet(2$^+$)+$\alpha$ channel is increased instead. 
The average of $\theta_\alpha^2$ roughly agrees between the experiment and theory,
however, the experimental 
$\theta_\alpha^2$ shows a larger spread between the 
resonances.
What is causing this discrepancy remains to be answered;
whether it is from the experimental resolution or a theoretical ambiguity, or 
physical properties of the resonance states. 
%The experimental uncertainty does not include the 
%the spectrum is not completely reproduced with the 
%R-matrix calculation, and 

There are factors that are not fully included in the calculation, such as 
1) the radial motion of $\alpha$ particle, 2) the rotational motion of 
\bet, and 3) the possible fragmentation of the state, coupling with 
other configurations.
The first factor may explain 
the larger theoretical $\theta_\alpha^2$ of the 2$^+$ and 4$^+$ resonances,
while the second factor is more relevant in the 0$^+$ resonance 
and the $\theta_\alpha^2$ may be underestimated.
It is also possible from the third factor that the $\theta_\alpha^2$ of the 0$^+$ resonance 
will be enhanced if the coupling of the \bet+$\alpha$ state with the continuum is
correctly included. 

From the experimental side, there are possible scenarios
in which $\theta_\alpha^2$  can deviate beyond the 
experimental uncertainty assigned in a standard manner.
%% which was evaluated only with
%% the single-channel R-matrix calculation with the minimum set of resonances.
One possibility is that the mixing ratio of the (5$^-$, 4$^+$) doublet
was not correctly determined, and the actual 4$^+$ component is stronger.
This is possible because of the limited orthogonality 
between those two resonances, and 
%which may not be very accurate in the current measurement, 
%The last line of Table~\ref{tab:resonance_params} shows 
%As an extreme case, we evaluated the $\theta_\alpha^2$ 
%necessary to reproduce the peak area only with a single 4$^+$ resonance. 
%We obtained %is unlikely because of the mismached resonant width,
%but the width can be considered 
we evaluate $\theta_\alpha^2(4^+)=7\%$ as a maximum limit by this effect.
%On the other hand, the dominance of 5$^-$ resonance is unlikely, 
%considering the observed angular distribution.
Another possibility is that 
the 4$^{+}$ resonance has a large neutron width $\Gamma_n$.
%which reduces the peak height and a larger $\Gamma_\alpha$ is required.
%The primary effect of introducing $\Gamma_n$ into the analysis is  
%a simple reduction of resonant height, and thus a larger $\Gamma_\alpha$ is required.
In that case,
$\Gamma_\alpha$ can be more than 100~keV ($\theta_\alpha^2(4^+) > 5\%$) when the resonance 
had a broad $\Gamma_n$ of over 300~keV.
There is no prediction of $\Gamma_n$ available for this resonance, but  
some of the 
neighboring 
resonances are reported to have $\Gamma_n$
of a similar order.
%The third possibility is for the broad $0^+$ resonance at 15.1 MeV. 
%We confirmed that introducing an additional nearby $0^+$ or $1^-$ state 
%with a smaller width
%will not cause an inconsistency to the 
%observed spectral shape or angular distribution.
%Then the $\theta_\alpha^2$ of the original $0^+$ resonance 
%can be significantly reduced.
%Thus, in all these cases, the agreement of experimental and theoretical $\theta_\alpha^2$ 
%could be improved, while the \jpi\ assignment is kept unchanged.

In summary,
we searched for resonances in \cft\ in the energy range \eex=14--19 MeV
with the resonant elastic scattering method
and found several $\alpha$-cluster-like states,  
obtaining new spectroscopic information % including \jpi,  
as displayed in Table~\ref{tab:resonance_params}.
In spite of many previous measurements with various methods,
the knowledge of observed resonances was quite limited, or completely absent.
%%the \jpi\ had been determined only for a limited number of states.
%We put a particular emphasis on the discovery of broad and low-spin resonances 
%around 15 MeV, which was not possible in previous measurements.
%This was made possible by the 
%$\alpha$-resonant elastic scattering with 
%the thick-target method in inverse kinematics.
%studied resonant states in \cft\ 
%$\alpha$-resonant elastic scattering with 
%the thick-target method in inverse kinematics,
%using a low-energy \bet\ beam at CRIB.
%and obtained the excitation functions of 
%\bet+$\alpha$ elastic scattering,
We put a special emphasis on the newly identified 3 resonances which exhibit 
level energy spacings and \jpi\ 
that perfectly 
agree with the prediction of a nuclear-cluster band of LCCS.
We claim this as the strongest indication of the LCCS ever found.
The comparison of the experimental and theoretical $\theta_\alpha^2$
is also performed in this work, %but with a limited precision.
and a rough agreement was observed between them. 
% in the partial widths, although
%the experimental widths appeared to be more deviated
%than the theoretical ones.
%The precision is still limited and 
%improvements are needed for both experimental and theoretical sides
%to perform a finer comparison of the widths, which 
A finer comparison may lead us to a 
more profound understanding of the LCCS.
As investigated in the theoretical calculation of the \cft\ system, 
the orthogonality between different quantum mechanical states 
is considered to play a key role in stabilizing the LCCS.
Further studies may reveal whether this mechanism 
is universal in nuclear systems or particular to \cft.
As an experimental technology, 
this achievement can be a milestone for the synthesis of 
nuclear cluster configurations
with more exotic topology,
such as triangles and rings \cite{Itagaki:04,Wilkinson:86}.

%\begin{acknowledgments}
We appreciate Prof. Y.~Kanada-En'yo for the discussion and suggestion 
on the theoretical implication.
The experiment was performed at RI Beam Factory operated
by RIKEN Nishina Center and CNS, the University of Tokyo.
We are grateful to the RIKEN and CNS accelerator staff for their beam production
and acceleration.
This work was partly supported by JSPS KAKENHI (No. 25800125, 15K17662, and 16K05369) in Japan,
and the National Research Foundation Grant funded by Korea Government 
(Grants Nos. NRF-2009-0093817, NRF-2016R1D1A1A09917463, NRF-2016R1A5A1013277, \\and 
NRF-2016K1A3A7A09005579).
This work was also supported in part by the Vietnam Academy of Science and Technology under the Program of Development in the field of Physics by 2020 - Study of unstable nuclei beam induced nuclear reactions at RIKEN.

%\end{acknowledgments}

\bibliography{crib}

\newpage

%\cite{Cappuzzello:15}

\begin{table*}[htbp]
\caption{\label{tab:resonance_params} 
The resonance parameters in \cft\ determined by the present work,
compared with the AMD calculation \cite{Suhara:10a}. 
Parameters in bold letters are for LCCS predicted in the calculation, and 
the corresponding experimental resonances.
Previously observed states
with their \jpi\ determined are also shown, but they do not 
necessarily correspond to the present measurement.   
See \cite{Oertzen:04,TUNL14:91,Resler:89,Soic:03,Milin:04,Price:06,Price:07,Haigh:08,Cappuzzello:15,Freer:14,Fritsch:16} for 
complete data, including other states.
Note that the theoretical \eex\ is after the threshold normalization.
}
\begin{tabular}{ccccccccc}
\hline\hline
\multicolumn{4}{c}{Present Work} & \multicolumn{3}{c}{Suhara \& En'yo \cite{Suhara:10a}} & \multicolumn{2}{c}{Other Experiments}\\
\eex\ (MeV) & $J^{\pi}$ & $\Gamma_{\alpha}$ (keV) & $\theta_{\alpha}^2$  & 
\eex\ (MeV) & $J^{\pi}$ & $\theta_{\alpha}^2$ & \eex\ (MeV) & $J^{\pi}$ \\
\hline
14.21        & (2$^{+}$)     &  17(5)         & 3.5\%      &     &     \\ 
14.50        & 1$^{-}$       &  45(14)        & 4.5\%      &     & && 14.67 &  6$^{+}$ \cite{Oertzen:04}       \\ %%%&     & && 14.63 &  (1$^{-}$) \cite{Resler:89}      \\   
             &               &                &            &     & && 14.717 &  4$^{+}$ \cite{Resler:89}       \\
             &               &                &            &     & && 14.87 &  5$^{-}$ \cite{Oertzen:04}       \\
{\bf 15.07} & {\bf 0$^{+}$} & {\bf 760(250)} & {\bf 34(12)\%} &{\bf 15.1} & {\bf 0$^{+}$}& {\bf 16\%}&15.20 &  4$^{-}$ \cite{Resler:89}       \\
             &               &                &            &     &       &    & 15.56 &  3$^{-}$ \cite{Price:07}       \\
{\bf 16.22} & {\bf 2$^{+}$} & {\bf 190(55)}  & {\bf 9.1(27)\%}&{\bf 16.0} & {\bf 2$^{+}$}& {\bf 15\%}& 15.91 &  4$^{+}$ \cite{Resler:89}      \\
16.37        & (4$^{+}$)     & 15(4)         &   3.0\% & && & 16.43  &  6$^{+}$ \cite{Oertzen:04}   \\
16.93        & (2$^{+}$)     & 270(85)       &  10.3\% & && & 16.9   & 0$^{+}$ \cite{Cappuzzello:15} \\
17.25        & (1$^{-}$)     & 190(45)       &  5.5\%  & && & 17.30  & 3$^{-}$ \cite{Freer:14}   \\
             &               &               &         & && & 17.30  & 4$^{-}$ \cite{Oertzen:04}   \\
             &               &               &         & && & 17.99  & 2$^{+}$ \cite{Freer:14}   \\
18.02        & (3$^{-}$)     & 31(19)        &  1.3\%  & && & 18.22  & 4$^{+}$ \cite{Freer:14}    \\
18.63        & 5$^{-}$       & 72(48)        &  9.4\%  & && & 18.83  & 5$^{-}$ \cite{Freer:14}   \\
{\bf 18.87}  & \bf{4$^{+}$}  & \bf{45(18)}&  {\bf 2.4(9)\%} &{\bf 19.2} & {\bf 4$^{+}$}& {\bf 9\%}&  \\
\hline\hline
%{\bf 18.72}\footnotemark[1]  & \bf{4$^{+}$}  & \bf{120(28)} &  {\bf 6.7(16)\%} & \\
\end{tabular}

\end{table*}

\end{document}